\documentclass[12pt]{article}
\usepackage{graphics}
\begin{document}
\begin{center}
{The Self-energy of Nucleon for the Pion-nucleon Scattering Parameter}
\end{center}
\begin{center}
{Susumu Kinpara}
\end{center}
\begin{center}
{\it Institute for Quantum Medical Science \\ Chiba 263-8555, Japan}
\end{center}
\begin{abstract}
The scattering amplitude of the pion-nucleon elastic scattering is calculated by the lowest-order approximation 
of the perturbative expansion with the non-perturbative term. 
The self-energy of nucleon is determined so as to give the scattering parameters of the $S$-wave.
It is shown that the scattering process mediated by the $\sigma$-meson plays a decisive role together with the pseudovector pion interaction.
\end{abstract}
%\vspace{4.mm}
\section*{\normalsize{1 \quad Introduction}}
%\vspace{4.mm}
\hspace*{4.mm}
The self-energy of nucleon at the zero density is expected to clarify the phenomena concerning to hadrons from the low energy to the high energy region.
The process is described by the field theoretical method of the meson-exchange model in which the fundamental elements are nucleon and meson. 
For pion is given the lightest mass and the longest range in comparison with the other heavior mesons it plays an important role to account for the nuclear system.
The interaction is stronger than the electromagnetic one and then we need the higher-order terms of the perturbative expansion to draw a meaningful result.
\\\hspace*{4.mm}
Besides the magnitude of the coupling constant the pion-nucleon interaction is thought to be the pseudovector coupling which contains the derivative 
and the usual procedure of the renormalization is not attainable.  
In place of the perturbative treatment the non-perturbative approach is useful to derive the general relation between the vertex part 
and the self-energy of the nucleon propagator as well as the Ward-Takahashi identity in the electromagnetic system.
The self-energy results in the finite quantity under the mass and the wave function renormalization by virtue of the non-perturbative term. 
At present our calculation is restricted to the lowest-order and the dependence on the coupling constant does not appear
because of the cancellation between the perturbative part and the non-perturbative part in the fraction. 
\\\hspace*{4.mm}
In order to calculate the scattering process containing the internal off-shell line of the nucleon propagator 
and compare with the experiment it is necessary to change the shape of the self-energy 
for each phenomenon and the incident energy separately. 
The models of the self-energy are classified by the maximum order of the expansion.
The type of the model suitable for the structure constant is selected individually by comparing the numerical results.
For example the constants of the electromagnetic interaction such as the anomalous magnetic moment of nucleon and the electromagnetic polarizability
of proton are accounted by one of these models respectively \cite{Kinpara}. 
\\\hspace*{4.mm}
On the pion-nucleon scattering it is probably not realistic to examine both sides of the resonance region by a single model.
The intermediate energy region is explained well by the model using the pion-nucleon-nucleon vertex with the non-perturbative term.
For the low energy region below the resonance energy any existing model of the self-energy of us does not give the correct answer to explain the experimental data. 
It implies that the spin-isospin structure of the partial wave is related to the higher-order processes or the heavier mesons which produce the interaction
and then the form of the self-energy is changed largely.
We attempt to search for the self-energy by which the low-energy parameters are reproduced.
\\
\section*{\normalsize{2 \quad The effect of the self-energy on the scattering parameters }}
%\vspace{4.mm}
\hspace*{4.mm}
There are two types of the pion-nucleon interaction available that is the pseudoscalar (PS) and the pseudovector (PV) couplings. 
In the present study the PV type is chosen to proceed the calculation of the amplitude of the pion-nucleon scattering
by the lowest-order approximation.
It is verified that these two interactions are connected by the non-perturbative term since the vertex part of the PV is expressed as follows
\begin{eqnarray}
\Gamma(p,q) = \gamma_5 \,\gamma\cdot (p-q) + G(p)^{-1} \,\gamma_5 + \gamma_5 \,G(q)^{-1}
\end{eqnarray}
where $p$ and $q$ are the outgoing and the incoming nucleons.
The pion field is the isovector with the mass $m$ and for the calculation of the diagram $\Gamma(p,q)$ is multiplied by the factor $-f/m \, \tau_i \;(i=1,2,3)$ 
on the Pauli matrix $\tau_i$.
The magnitude of the coupling constant $f$ of the PV interaction is known to be $f \sim 1$ by the proton-neutron elastic scattering
and the pion-nucleon scattering. 
At first the empirical value is used herein although the inclusion of the non-perturbative term tends to reduce it
according to our results of the calculation of the magnetic moment of nucleon and the pion-nucleon scattering at the intermediate energy region.
\\\hspace{4.mm}
When either of the momentum $p$ or $q$ is off-shell the non-perturbative term survives as seen in Eq. (1).
Making the assumption of the self-energy $\Sigma(p)$ to exist the nucleon propagator $G(p)\,=\,(\gamma\cdot p-M-\Sigma(p))^{-1}$ is responsible for
changing the form of the vertex in comparison with that of the PS interaction.
Because of the non-perturbative term the renormalization procedure works. 
The model of the self-energy is effective against various phenomena of the electromagnetic interaction
in which the photon energies are associated with the region from the low to the intermediate energy.
\\\hspace{4.mm}
When we try to account for the scattering parameters the application of the models to the low energy pion-nucleon scattering encounters difficulties.
Then the appropriate form is to be determined by considering the experimental data.
The general form of $\Sigma(p)$ is given in terms of the coefficients $c_i(p^2)\;(i=1,2)$ as   
\begin{eqnarray}
\Sigma(p) = M c_1(p^2) - \gamma\cdot p \, c_2(p^2)
\end{eqnarray}
We pay attention to small area of the on-shell condition $p^2 = M^2$ such as $p^2-M^2 \sim m M \sim 0.1$ [GeV]${}^{2}$ 
and the expansion of $c_i(p^2)$ in the series of $p^2-M^2$ is assumed to be valid
\begin{eqnarray}
c_i(p^2) = \sum_{n=0}^{\infty} \frac{1}{n!} \, c^{(n)}_i (p^2-M^2)^n
\end{eqnarray}
The condition of the mass and the wave function renormalization makes $\Sigma(p)$ start from the quadratic order in $\gamma\cdot p-M$.
It is expressed by the relations on the coefficients of the expansion in Eq. (3) as
\begin{eqnarray}
c^{(0)}_1 = c^{(0)}_2
\end{eqnarray}
\begin{eqnarray}
c^{(1)}_1 - c^{(1)}_2 = \frac{c^{(0)}_2}{2 M^2} 
\end{eqnarray}
The relation in Eq. (5) is equivalent to the Ward identity for nucleon.
There is not any constraint about the coefficients of the higher-order $c^{(n)}_i\;(n\ge 2)$
as far as the framework of the one photon vertex.
\\\hspace{4.mm}
For the calculation of the diagram the rationalized form of the propagator is useful.
Making use of the $c_i(p^2)$ mentioned above the renormalized propagator is given as
\begin{eqnarray}
G(p) = \frac{\alpha(p^2) \gamma \cdot p +\beta(p^2) M}{p^2-M^2}
\end{eqnarray}
\begin{eqnarray}
\alpha(p^2)=(1+c_2(p^2))^{-1}\frac{p^2-M^2}{p^2-(M+\hat{M}(p^2))^2}
\end{eqnarray}
\begin{eqnarray}
\beta(p^2)=\alpha(p^2)(1+\frac{\hat{M}(p^2)}{M})
\end{eqnarray}
\begin{eqnarray}
\hat{M}(p^2) = M \, \frac{c_1(p^2)-c_2(p^2)}{1+c_2(p^2)}
\end{eqnarray}
Applying $G(p)$ to the scattering amplitude the $\alpha(p^2)$ and $\beta(p^2)$ are expanded in powers of $p^2-M^2$ like $c_i(p^2)$
\begin{eqnarray}
\alpha(p^2) = \sum_{n=0}^{\infty} \frac{1}{n!} \, \alpha^{(n)} (p^2-M^2)^n
\end{eqnarray}
The $\beta(p^2)$ is analogous to Eq. (10).
The renormalization conditions in Eqs. (4) and (5) are converted to the relation
$\alpha^{(0)} = \beta^{(0)} = 1$.
\\\hspace{4.mm}
The relation of the higher-order is obtainable.
For example the coefficients of the next order are shown explicitly
\begin{eqnarray}
\qquad \qquad \quad \alpha^{(1)} = \frac{1}{4 M^2}\cdot \frac{c^2}{1+c}-c^{(1)}_2   \qquad\qquad (c \equiv c^{(0)}_2)
\end{eqnarray}
\begin{eqnarray}
\beta^{(1)} = \frac{1}{2 M^2}\cdot \frac{c}{1+c}(1+\frac{c}{2})-c^{(1)}_2
\end{eqnarray}
up to the $c_i^{(1)}$ order in $c_i(p^2)$.
The approximated ones $\alpha^{(1)}$ and $\beta^{(1)}$ are used in the present study. 
It is noted that when $c$ is staying near the value $c=-1$ the procedure of the expansion may not work.
In that case the correction of the renormalized propagator would enlarge the contribution of the intermediate state too much.
\\\hspace{4.mm}
The low-energy pion-nucleon scattering is determined by the invariant matrix element $T$.
It is decomposed by the isospin and the gamma matrices on account of the symmetry in terms of the coefficients $A^{(\pm)}$ and $B^{(\pm)}$
\begin{eqnarray}
T =  T^{(+)}-\vec{\tau}\cdot \vec{t} \;\, T^{(-)}
\end{eqnarray}
\begin{eqnarray}
T^{(\pm)} = A^{(\pm)} + \frac{(q+q^\prime) \cdot \gamma}{2} B^{(\pm)}
\end{eqnarray}
where $(t_c)_{ba}\equiv -i \epsilon_{cba}$ with $\epsilon_{123}=1$.
The $q$ and $q^\prime$ are the four-momenta of the incident and the scattered pion accordingly.
\\\hspace{4.mm}
The lowest-order process of the perturbative expansion is calculated with the use of the vertex (Eq. (1)) and the propagator (Eq. (6)).
The improved form of the interaction has an effect on the coefficients $A^{(\pm)}$ and $B^{(\pm)}$ and it is given as follows
\begin{eqnarray}
&&A^{(\pm)}(s,t,u) = -\frac{1 \pm 1}{2}\frac{g^2}{M} 
+\frac{g^2}{4 M}[\,\alpha(s)+\beta(s)\pm\alpha(u)\pm\beta(u)\nonumber\\\nonumber\\
&& \qquad\qquad\qquad\qquad -c_1(s)-c_2(s)\mp c_1(u)\mp c_2(u)\,]
\end{eqnarray}
\begin{eqnarray}
&&B^{(\pm)}(s,t,u) = \frac{1 \mp 1}{2} \frac{g^2}{2 M^2} + \frac{g^2}{2} [\,\frac{-\alpha(s) \pm\alpha(u) -c_2(s) \pm c_2(u)}{2 M^2}\nonumber\\\nonumber\\
&& \qquad\qquad\qquad\qquad -\frac{\alpha(s)+\beta(s)}{s-M^2} \pm \frac{\alpha(u)+\beta(u)}{u-M^2}\,]
\end{eqnarray}
where $s = (p+q)^2$, $t = (q-q^\prime)^2$ and $u = (p-q^\prime)^2$ are the Mandelstam variables 
with the four-momenta of the initial nucleon $p$ and the final nucleon $p^\prime$.
These variables are connected by the conservation of momentum $p+q = p^\prime +q^\prime$.
As has been stated the connection between PV and PS is verified by setting the self-energy equal to zero.
The quantities $A^{(\pm)}(s,t,u)$ and $B^{(\pm)}(s,t,u)$ result in those of the PS interaction \cite{Serot} with the coupling constant
$g \equiv 2 M f /m$ by the approximation.
\\\hspace{4.mm}
According to the method of the partial wave expansion 
the scattering amplitude $f_{l\pm}(w) = {\rm exp}[i \delta_{l \pm}] \,{\rm sin} \,\delta_{l \pm}/\vert\vec{q}\vert$ plays a decisive role 
in calculating the pion-nucleon scattering.
The explicit form of $f_{l\pm}(w)$ as a function of the center of mass total energy $w \equiv \sqrt{\vec{q}^{\; 2}+M^2}+\sqrt{\vec{q}^{\; 2}+m^2}$ 
is seen in Ref. \cite{Walecka}.
It is expanded in the series of the square of the pion momentum $\vec{q}^{\; 2}$ up to the $O(\vec{q}^{\; 2})$ order as
\begin{eqnarray}
f_{0+}(w) = a_{0} + r_{0} \,\vec{q}^{\; 2} +O(\vec{q}^{\; 4})
\end{eqnarray}
\begin{eqnarray}
\qquad \quad f_{l\pm}(w) = a_{l\pm} \,(\vec{q}^{\; 2})^l +O((\vec{q}^{\; 2})^{l+1}) \qquad (l \ge 1)
\end{eqnarray}
for each component classified by the orbital angular momentum $l$ and the total angular momentum $j=l \pm 1/2$.
In the present study eight coefficients $a_0^{(\pm)}$, $r_0^{(\pm)}$ and $a_{1 \pm}^{(\pm)}$ are taken to examine the effect of the self-energy
at the low energy region.
The superscripts ($\pm$) of them correspond to those of $A^{(\pm)}$ and $B^{(\pm)}$ specifying the isospin degrees of freedom.
\\
\section*{\normalsize{3 \quad The calculation of the scattering parameters}}
%\vspace{4.mm}
\hspace*{4.mm}
The correction of the scattering matrix for the lowest-order process is twofold.
First addition of the non-perturbative term changes the form of the vertex and second the self-energy of the propagator
has an effect on the internal line of the diagram.
In order to examine the function form appropriate to the scattering parameters the $c_i(p^2)$ are expanded as shown in Eq. (3).
The equation with respect to the constant values $c_i^{(n)}$ is solved by the method of the matrix inversion.
We approximate the nucleon propagator as $\alpha(v) \rightarrow 1$ and $\beta(v) \rightarrow 1$ irrespective of $v = s \;{\rm or} \;u$
because the expansion of $\alpha(v)$ and $\beta(v)$ prevents the linearization of the relations on $c_i^{(n)}$ as seen 
from the denominators in Eqs. (11) and (12).
\\\hspace*{4.mm}
The scattering parameters of the $S$-wave ($l$=0) are $a_0^{(\pm)}$ and $r_0^{(\pm)}$ up to $O(\vec{q}^{\; 2})$ 
and four unknown coefficients $c\;(\equiv c_1^{(0)} = c_2^{(0)})$, $c_2^{(1)}$, $c_1^{(2)}$ and $c_2^{(2)}$ are determined by solving the linear equation.
The other one $c_1^{(1)}$ is dependent variable on $c$ and $c_2^{(1)}$.
These relations are expressed by the matrix form as
\begin{eqnarray}
{\rm Y} = {\rm V} + {\rm \Delta} \, {\rm X}
\end{eqnarray}
in which $Y=(m a_0^{(\pm)},\, m^3 r_0^{(\pm)})$ and $X=(c,\, M^2 c_2^{(1)},\, M^4 c_1^{(2)},\, M^4 c_2^{(2)})$
are the dimensionless vectors.
The vector $V$ is a part which consists of the components independent of the self-energy.
The matrix inversion is performed and the solution $X$ is obtained 
\begin{eqnarray}
{\rm X} = {\rm \Delta}^{-1} \, ({\rm Y}-{\rm V})
\end{eqnarray}
under the condition that ${\rm det \, \Delta} \neq 0$. 
The respective values of four components in $Y$ are taken from the experimental values of the $S$-wave parameters listed in Table 1.
For the input data the choice of the other set is possible and changes the output values of $X$.
It has been verified that replacing one of the parameters $r_0^{(-)}$ from $0.007\pm 0.005$ \cite{Pilkuhn} to $-0.013\pm 0.006$ \cite{Dunbrajs} does not vary
the outputs much and parameters of the $P$-wave are shifted only about 10$\%$ at most.
\\\hspace*{4.mm}
The existence of the solution is not influenced by the detail of $V$.
It makes us possible to add any favorable process which is independent of the self-energy.
For example the pion-nucleon interaction is generated by the exchange of the isoscalar-scalar boson which is assumed to be the $\sigma$-meson \cite{Serot}.
It is introduced to construct the scattering amplitude.
In the present study the interaction of the $\sigma$-meson exchange is also applied
using the parameters of the coupling constants of the N-N-$\sigma$ and the $\sigma$-$\pi$-$\pi$ virtices and the $\sigma$-meson mass.
\\\hspace*{4.mm}
The matrix inversion determines the solution $X$ so as to give the low-energy part of the $S$-wave completely.
It yields $c = -0.39$, $c_2^{(1)} = 0.28$, $c_1^{(2)} = -8.39$ and $c_2^{(2)} = 5.59$ on the coefficients of the self-energy under the $\sigma$-meson interaction.
One interesting point is that the numerical value of $c$ is not in the vicinity of $c = -1$.
The result is contrary to the expectation of the pion-nucleon-nucleon three-point vertex 
that it approaches to the original form of the PV interaction ($\,\Gamma(p,q)\,\sim\,\gamma_5 \gamma \cdot (p-q)\,$) 
in other words the non-perturbative term disappears.
Then it enables us apply the expansion of $\alpha(p^2)$ 
to use the approximated form ($\alpha(p^2) \approx 1 + \alpha^{(1)} (p^2-M^2)$) for the calculation of the scattering parameters.
\\\hspace*{4.mm}
Next our interest is to evaluate the $P$-wave scattering paramers.
The procedure of the expansion of four elements $f_{1\pm}^{(\pm)}$ is same as the $l=0$ case 
and the parameters are obtained by $a_{1\pm}^{(\pm)} = \lim_{\vec{q}^{\,2} \rightarrow 0}\,d f_{1\pm}^{(\pm)} /d \vec{q}^{\,\,2}$.
The experimental values of these four parameters help us to search for the spin and isospin dependence of the interaction and the non-perturbative effect.
As an example the shift of $\vert a_{1+}^{\, (-)}/a_{1+}^{\, (+)}\vert$ from 1 is an indicator of the strength of the isospin dependent part.
It has turned out that the inclusion of the self-energy is not a useful way since the result of the calculation does not produce the shift.
The preceding additional force of the $\sigma$-meson is required to correct the volume suitably by the $t$-dependent term.
The coupling constant of the N-N-$\sigma$ vertex is related to the nuclear many-body system and the value is assumed to be roughly $g_{\sigma} \sim 8$.
The strength of the $\sigma$-$\pi$-$\pi$ coupling constant $g_{\sigma\pi\pi}$ is prepared taking into account the effect of the self-energy.
\\\hspace*{4.mm}
The scattering parameters are shown in the nuclear unit which makes the quantities dimensionless.
The $no\;\sigma$ model is the calculation without the $\sigma$-meson exchange interaction.
On the other hand the $with\;\sigma$ model is the calculation with the $\sigma$-meson exchange interaction.
The numerical values of the parameters in the present calculation 
are $M$=\,0.939$\,$[GeV], $m$=\,0.138$\,$[GeV], $f$=1, $m_\sigma$=$\,$0.52$\,$[GeV], $g_\sigma=\,$8.36 and $g_{\sigma\pi\pi}=\,$7.
These are the standard values associated with the calculation of the nuclear structure and the nucleon-nucleon elastic scattering. 
Determining the magnitude of $g_{\sigma\pi\pi}$ the experimental value of the isospin $I$=3/2 part of the amplitude $a_{1+}^{(+)}-a_{1+}^{(-)}$ has been used 
neglecting the shift from the free propagator when the matrix inversion is performed.
The experimantal data are taken from Ref. \cite{Serot}.
They are also seen in Ref. \cite{Pilkuhn}.
\begin{center}
\begin{tabular}{|c|c c c c c c c c|}
   \multicolumn{9}{c}{ Table 1 }\\
      \hline
 & $a_0^{(+)}$ & $a_0^{(-)}$ & $r_0^{(+)}$ & $r_0^{(-)}$ & $a_{1-}^{(+)}$ & $a_{1-}^{(-)}$ & $a_{1+}^{(+)}$ & $a_{1+}^{(-)}$ \\
      \hline
no $\sigma$ & {\small 0.325} & {\small 0.124} & {\small 0.32} & {\small 0.043} & {\small -0.055} & {\small -0.014} & {\small 0.077} & {\small -0.077} \\
     \hline
with $\sigma$ & {\small -0.022} & {\small 0.098} & {\small -0.06} & {\small 0.006} & {\small -0.044} & {\small -0.036} & {\small 0.116} & {\small -0.066} \\
     \hline
Exp.  & {\small -0.015} & {\small 0.097} & {\small -0.06} & {\small 0.007} & {\small -0.056} & {\small -0.013} & {\small 0.133} & {\small -0.081} \\
 &{\footnotesize$\pm$0.015}&{\footnotesize+0.003}&{\footnotesize$\pm$0.02}&{\footnotesize$\pm$0.005}
&{\footnotesize$\pm$0.010}&{\footnotesize$\pm$0.003}&{\footnotesize$\pm$0.003}&{\footnotesize$\pm$0.002}\\
     &           &{\footnotesize -0.007}& & & & & & \\
     \hline
\end{tabular}
\end{center}
\vspace{2.mm}
\hspace*{4.mm}
The meaning of the $\sigma$-meson in the low energy region is made clear by increasing the value of $g_{\sigma\pi\pi}$.  
Making use of $g_{\sigma\pi\pi}=\,$12 the same calculation results in $c = 0.07$, $c_2^{(1)} = 0.02$, $c_1^{(2)} = -13.4$ and $c_2^{(2)} = 8.05$.
The $c$ and $c_2^{(1)}$ are much smaller than those of $g_{\sigma\pi\pi}=\,$7 and the $c_i^{(2)}$ values are not used for the propagator
in the calculation of the scattering parameters. 
Consequently the approximation by the free propagator is valid to give the $S$-wave parameters 
in which the deviation from the starting values is not large.
It implies that the $\sigma$ interaction plays a role similar to the self-energy in the lowest-order process.
There would be some relations between these two terms.
\\\hspace*{4.mm}
When the internal fermion line of nucleon is approximated by the free propagator ($\alpha(p^2) = \beta(p^2) = 1$) 
the $S$-wave components are certainly in agreement with the experimental values by definition.
The difference from it is ascribed to the higher-order correction due to $\alpha^{(1)}$ and $\beta^{(1)}$ at the low-energy limit. 
In the $no\,\sigma$ model the shift of the $S$-wave from the experimental value is more obvious than that of the $with\,\sigma$ model.
It is explainable by the function form of the self-energy in each model. 
For the $no\,\sigma$ model they are $c = -1.03$, $c_2^{(1)} = 0.64$, $c_1^{(2)} = -1.32$ and $c_2^{(2)} = 2.14$ 
and then the $\alpha^{(1)}$ and $\beta^{(1)}$ ($\,\sim (1+c)^{-1}\,$) increase largely, so that the signs of $a_0^{(+)}$ and $r_0^{(+)}$ are reversed.
The next order $\alpha^{(2)}$ and $\beta^{(2)}$ are required to verify the convergence of the expansion and make a conclusion on the results.
\\\hspace*{4.mm}
The feature of the $no\,\sigma$ model is seen in the higher component of the partial wave too. 
The invariant amplitudes of the $D$-wave are
given by the linear combinations as $a^{1/2} = a^{(+)}+2 a^{(-)}$ and $a^{3/2} = a^{(+)}- a^{(-)}$ for the isospin $I$=1/2 and 3/2 state.
In Table 2 the data of the experiment is taken from Ref. \cite{Koch}.
These are in the nuclear unit same as Table 1.
The minus sign in the $a_{2-}^{1/2}$ component is the consequence of the excess of $\alpha^{(1)}$ and $\beta^{(1)}$ 
to compensate for the $\sigma$-meson exchange process.
\vspace{2.mm}
\begin{center}
\begin{tabular}{|c|c c c c|}
   \multicolumn{5}{c}{ Table 2 }\\
      \hline
 & $10^2\times a_{2-}^{1/2}$ & $10^2\times a_{2-}^{3/2}$ & $10^2\times a_{2+}^{1/2}$ & $10^2\times a_{2+}^{3/2}$ \\
      \hline
no $\sigma$ & {\small $-$0.14} & {\small 0.29} & {\small 0.34} & {\small $-$0.69} \\
     \hline
with $\sigma$ & {\small 0.15} & {\small 0.46} & {\small 0.61} & {\small $-$0.37} \\
     \hline
Exp.  & {\small 0.24} & {\small 0.21} & {\small 0.46} & {\small $-$0.50} \\
 &{\footnotesize$\pm$0.04}&{\footnotesize$\pm$0.02}&{\footnotesize$\pm$0.03}&{\footnotesize$\pm$0.02} \\
     \hline
\end{tabular}
\end{center}
\vspace{2.mm}
\section*{\normalsize{4 \quad Summary and remarks }}
%\vspace{4.mm}
\hspace*{4.mm}
With respect to the method of the self-energy it is difficult to explain the behavior of the low energy region below the resonance on an equal footing 
with the counterpart of the intermediate energy for the pion-nucleon scattering.  
Consequently a rather practical way has been employed to obtain the low energy parameters starting from the PV coupling interaction with the non-perturbative term.
The second-order term in the expansion of the self-energy has been included to construct the propagator.
Numerically the exact treatment makes the results worse and which means that the linearization should be improved to solve the matrix equation.
In addition to the effect of the self-energy the process mediated by the $\sigma$-meson is essential to understand the elastic scattering at the low energy region.
It may be attributed to the $\pi$-$\pi$-N-N vertex combining the internal nucleon propagator with the external pions
as well as the case of the compton scattering particularly in the forward direction.
\\
%\section*{\normalsize{ Appendix }}
%
\hspace{4.mm}

\end{document}